# Weighted Secrecy Coverage Analysis and the Impact of Friendly Jamming over UAV-Enabled Networks

X. A. Flores Cabezas; D. P. Moya Osorio; and M. Latva-aho
Centre for Wireless Communications, University of Oulu, Finland
Email: [xavier.florescabezas, diana.moyaosorio, matti.latva-aho]@oulu.fi

***Abstract*—In 5G and beyond networks, Unmanned Aerial Vehicles (UAV) are an attractive solution to enhance the secrecy of a wireless systems by exploiting their predominant LOS links and spacial manoeuvrability to introduce a friendly jamming. In this work, we investigate the impact of two cooperative UAV-based jammers on the secrecy performance of a ground wireless wiretap channel by considering secrecy-area related metrics, the jamming coverage and jamming efficiency. Moreover, we propose a hybrid metric, the so-called Weighted Secrecy Coverage (WSC) that can be used as a metric for gaining insights on the optimal deployments of the UAV jammers to provide the best exploration of jamming signals. For evaluating these metrics, we derive a closed-form position-based metric, the secrecy improvement, and propose an analogous computationally simpler metric. Our simulations show that a balanced power allocation between the two UAVs leads to the best performances, as well as a symmetrical positioning behind the line of sight between the legitimate transmitter and receiver. Moreover, there exist an optimal UAV height for the jammers. Finally, we propose a sub-optimal and simpler problem for the maximisation of the WSC.**

## I. Introduction

The fifth generation wireless networks (5G) will not be able to fulfil the requirements of the future in 2030 and beyond, thus the sixth generation (6G) wireless communication networks are expected to finally provide the requirements of wireless ubiquity, new intelligent radio design, increased energy and spectral efficiency, high level of intelligence and automation, and robust security mechanisms [1]. To provide wireless ubiquity, 6G should enhance network coverage beyond the terrestrial domain; for instance, by incorporating satellite and unmanned aerial vehicle (UAV) communications.

UAVs can potentially facilitate new applications and services by offering several advantages, such as on-demand coverage, dynamic and cost-effective deployment, fast response to service demands, mobility in three-dimensional (3D) space, and short-distance Line-of-Sight (LoS) links that can exploit millimeter waves to boost capacity and improve quality of service (QoS). Particularly, the use of UAVs is interesting in a wide range of 5G and beyond use-cases, such as, intelligent agriculture, environmental monitoring, infrastructure inspection, and disaster prevention and management [2], [3]. UAVs can be integrated to 5G/6G infrastructure as aerial base stations, aerial user equipments or flying relays.

At the same time that UAV communications offer numerous benefits, Air-to-Ground (A2G) communications are also prone to security breaches if they are misused by unauthorized agents for malicious purposes [4]. Indeed, several security challenges are expected due to the broadcasting nature of the wireless medium. Thus, information security is one of the fundamental requirements in this kind of systems [5]. Moreover, malicious aerial nodes can take advantage of their high mobility and flexibility to track their targets over time, thus overhearing or jamming their communications more effectively. On the other hand, the use of UAVs can also be attractive due to its potential to improve the security of wireless communications by exploiting their flexible deployments and dominant LoS links comparing to ground nodes.

Recently, due to the challenges to provide security provisioning in UAV-based networks with several constraints, physical layer security techniques have drawn special attention in a number of works as a compelling option for traditional upper-layer cryptographic mechanisms [6]–[9]. For instance, in [7], the outage probability (OP) of a legitimate receiver and the intercept probability (IP) of an eavesdropper are evaluated under the presence of a UAV-based friendly jammer. Therein, a metric so-called intercept probability security region is proposed and maximised by optimising the 3D deployment and power of the UAV jammer, given an OP constraint. In [8], the secrecy outage probability and the average secrecy rate are analyzed for a UAV-based system, where a ground IoT device transmits confidential information to a UAV, while a random number of ground randomly located eavesdroppers try to leak information. The analysis is carried out for the case without and with a friendly UAV that generates jamming signals to distract the eavesdroppers. In [9], the authors consider a confidential communication between a pair of ground legitimate nodes assisted by an energy-constrained UAV-enabled relay in the presence of multiple eavesdroppers with imperfect locations. In that system, simultaneous information and power transfer is considered to charge the UAV, and the legitimate destination works in full-duplex mode to transmit a friendly jamming signal while receiving the confidential signal in order to enhance the secrecy performance. The worst case secrecy rate maximisation problem is tackled by jointly optimising the position of the UAV-enabled relay, the power of the jamming signal, and the charging parameters. It was demonstrated that the jamming-based joint design achieves a significant improvement in the min-max secrecy rate.

The objective of this work is to evaluate the impact of UAV-enabled friendly cooperative jamming on the secrecy of a ground wireless wire-tap channel having no information about the eavesdropper's channel, inspired by the secrecy-related area performance metrics proposed in [10]. Particularly, we evaluate the area-based metrics, jamming coverage and jamming efficiency, and propose the Weighted Secrecy Coverage (WSC) metric as a helpful tool to evaluate the impact of friendly jamming on the enhancement of secrecy in UAV-enabled networks. We also formulate a sub-optimal optimisation problem to get insights on the maximisation of the WSC over the angle between the jammers in terms of, among other variables, the distance between Alice and Bob. To this end, we also compute the jamming improvement metric and propose an analogous variation that simplifies the computations and retains the overall behaviour, which is used in a proposed hybrid secrecy area metric, the Weighted Secrecy Coverage.

The rest of the paper is organised as follows: Section II provides an overview of the model of the system in question, explaining its parameters, constraints and fleshes out the propagation formulas for the different links. Section III details the secrecy metrics to be used from the secrecy capacity of the wiretap channel and culminating in our main secrecy area metric, the proposed Weighted Secrecy Coverage. In Section IV we run a number of simulations over various degrees of freedom to further analyse their impact and patterns that would allow us to assume a reduced number of degrees of freedom while maintaining optimality in our performance metric. In Section IV-A we propose an analogous jamming improvement function that simplifies computations and retains the behaviour of our main metric and lay out the optimisation problem by proposing a sub-optimal function to minimise, which approximates the Weighted Secrecy Coverage metric. Section V presents the conclusions and Appendix A fleshes out the secrecy improvement metric derivation.

## II. System Model

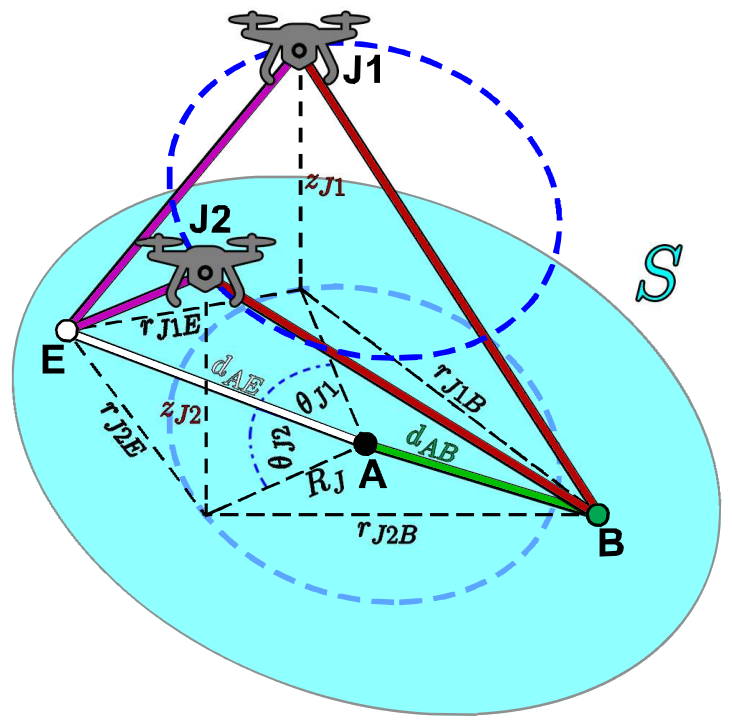

Fig. 1. Physical topology of constrained wiretap channel model with friendly jammers.

Consider the system depicted in Fig. 1 composed by a legitimate ground transmitter, Alice (A) at position $(x_{\mathrm{A}}, y_{\mathrm{A}})$; a legitimate ground receiver, Bob (B) at position $(x_{\mathrm{B}}, y_{\mathrm{B}})$; a ground eavesdropper, Eve (E) at position $(x_E, y_E)$; and two UAV-enabled friendly jammers, $\mathrm{J}_i$ with $i \in \{1,2\}$ at positions $(x_{\mathrm{J1}}, y_{\mathrm{J1}})$ and $(x_{\mathrm{J2}}, y_{\mathrm{J2}})$ respectively. In this scenario, we consider that E is located at an arbitrary position of a circular space, $S$, and tries to leak confidential information from the communication between A and B. The ground links are assumed to undergo Rayleigh fading and are subject to additive white Gaussian noise (AWGN), with mean power $\sigma^2$. Therefore, the corresponding channel coefficients, $h_{\mathrm{AU}}$ with $\mathrm{U} \in \{\mathrm{B}, \mathrm{E}\}$, are independent complex circularly-symmetric Gaussian random variables with variance $\mathbb{E}\left[|h_{\mathrm{AU}}|^2\right] = d_{\mathrm{AU}}^{-\alpha}$, where $d_{\mathrm{AU}}$ being the distance among a pair of nodes and $\alpha$ being the pathloss exponent for the ground links. Thus, the channel gains, $|h_{\mathrm{AU}}|^2$, follow an exponential distribution with parameter $\Omega_{\mathrm{AU}} = d_{\mathrm{AU}}^{\alpha}$. To enhance the secrecy performance of the system, the two friendly jammers are positioned so that they introduce artificial pseudo-noise in order to interfere E. Herein, we assume that the aerial jammers move in a three-dimensional space while their projection into the ground plane is confined to a circumference of radius $R_{\mathrm{J}}$ around A, so we express their positions in cylindrical coordinates $(r_{\mathrm{J1}}, \theta_{\mathrm{J1}}, z_{\mathrm{J1}})$ and $(r_{\mathrm{J2}}, \theta_{\mathrm{J2}}, z_{\mathrm{J2}})$ respectively. The A2G links, between the UAV jammers and B or E, can experience either line-of-sight (LoS) or non-line-of-sight (NLoS) propagation [11]. Therefore, the average pathloss for the jamming links are given by

$$\begin{aligned} L_{\mathrm{J}_i\mathrm{U}} &= P_{\mathrm{LoS}}\eta_{\mathrm{LoS}} d_{\mathrm{J}_i\mathrm{U}}^{\alpha_{\mathrm{J}}} + P_{\mathrm{NLoS}}\eta_{\mathrm{NLoS}} d_{\mathrm{J}_i\mathrm{U}}^{\alpha_{\mathrm{J}}}, \\ &= \left(z_{\mathrm{J}_i}^2 + r_{\mathrm{J}_i\mathrm{U}}^2\right)^{\frac{\alpha_{\mathrm{J}}}{2}} \left(P_{\mathrm{LoS}}\eta_{\mathrm{LoS}} + P_{\mathrm{NLoS}}\eta_{\mathrm{NLoS}}\right), \end{aligned} \tag{1}$$

where $z_{\mathrm{J}_i}$ is the height of $\mathrm{J}_i$, $r_{\mathrm{J}_i\mathrm{U}}$ is the distance from node U and the projection on the plane of $\mathrm{J}_i$, $\alpha_{\mathrm{J}}$ is the pathloss exponent for the A2G links, $\eta_{\mathrm{LoS}}$ and $\eta_{\mathrm{NLoS}}$ are the attenuation factors for the LoS and the NLoS links, respectively. Also, $P_{\mathrm{LoS}}$ and $P_{\mathrm{NLoS}}$ are the probabilities of LOS and NLOS connection, respectively given by

$$P_{\mathrm{LoS}} = \frac{1}{1 + \psi \exp\left(-\omega\left[\frac{180}{\pi}\tan^{-1}\left(\frac{z_{\mathrm{J}_i}}{r_{\mathrm{J}_i\mathrm{U}}}\right) - \psi\right]\right)}, \tag{2}$$

$$P_{\mathrm{NLoS}} = 1 - P_{\mathrm{LoS}}, \tag{3}$$

where $\psi$ and $\omega$ are environmental constants [12], [13].

Herein, we will evaluate the impact of the friendly jamming signal over the secrecy performance of the proposed system in terms of two secrecy metrics proposed in [10], the jamming Coverage and the jamming Efficiency, which will be described next. In that case, we are interested on analysing the gain obtained by the friendly jammer. Therefore, the received signal-to-noise ratios (SNRs) at B and E, are given by $\gamma_{\mathrm{B}} = a_j \cdot |h_{\mathrm{AB}}|^2, \gamma_{\mathrm{E}} = b_j \cdot |h_{\mathrm{AE}}|^2$, with $j \in \{\mathrm{NJ}, \mathrm{J}\}$, for the scenarios with and without friendly jamming, respectively. Thus, $a_{\mathrm{NJ}} = b_{\mathrm{NJ}} = \gamma_{\mathrm{A}}$, with $\gamma_A$ and $\gamma_{\mathrm{J}_i}$ being the transmit SNRs at A and $\mathrm{J}_i$, with $\gamma_{\mathrm{A}} = \frac{P_{\mathrm{A}}}{\sigma^2}$ and $\gamma_{\mathrm{J}_i} = \frac{P_{\mathrm{J}_i}}{\sigma^2}$, and $P_{\mathrm{A}}$ and $P_{\mathrm{J}_i}$ are the transmit powers at Alice and the jamming power at $\mathrm{J}_i$, respectively. Accordingly, for the friendly jamming case, $a_{\mathrm{J}}$ and $b_{\mathrm{J}}$ are given by

$$a_{\mathrm{J}} = \frac{\gamma_A}{1 + \frac{\gamma_{\mathrm{J}_1}}{L_{\mathrm{J}_1\mathrm{B}}} + \frac{\gamma_{\mathrm{J}_2}}{L_{\mathrm{J}_2\mathrm{B}}}}, \tag{4}$$

$$b_{\mathrm{J}} = \frac{\gamma_A}{1 + \frac{\gamma_{\mathrm{J1}}}{L_{\mathrm{J}_1\mathrm{E}}} + \frac{\gamma_{\mathrm{J2}}}{L_{\mathrm{J}_2\mathrm{E}}}}. \tag{5}$$

## III. Performance Analysis

In this Section, we focus our analysis on secrecy metrics that allow us to evaluate the impact of friendly jamming on the secrecy performance of the proposed system, while considering the fact that the location of E is unknown, but within the working area $S$. For this purpose, we resort to the secrecy metrics proposed in [10], jamming Coverage and jamming Efficiency, and we propose a hybrid metric, so-called Weighted Secrecy Coverage.

These metrics are expressed in terms of the well-known secrecy outage probability (SOP), which is the probability that the secrecy capacity falls below a certain secrecy rate $R_S$ [14], given by

$$\mathrm{SOP} = \Pr\left[C_S < R_S\right], \tag{6}$$

where $C_S$ is the secrecy capacity defined as the maximum achievable rate to maintain a confidential communication [14], that is, the maximum value between 0 and the difference between the capacities of the legitimate channel $C_{\mathrm{B}}$ and the wiretap channel $C_{\mathrm{E}}$ given by [14]

$$C_S = \left[C_{\mathrm{B}} - C_{\mathrm{E}}\right]^+ = \left[\log_2\left(\frac{1+\gamma_{\mathrm{B}}}{1+\gamma_{\mathrm{E}}}\right)\right]^+, \tag{7}$$

with $[X]^+ = \max[X, 0]$. Then, we define $\Delta$ as the secrecy improvement due to friendly jamming given as the ratio of the attained SOP without considering friendly jamming and the attained SOP with friendly jamming, and it is expressed as [10]

$$\Delta = \frac{\mathrm{SOP}_{\mathrm{NJ}}}{\mathrm{SOP}_{\mathrm{J}}}. \tag{8}$$

From (8), we can notice that $\Delta$ provides information of the impact of jamming at a certain position for Eve. In this reason, for the positions where $\Delta < 1$, the presence of jamming worsens the secrecy performance; whereas, if $\Delta > 1$, the presence of jamming improves the secrecy performance. Analysing this for every point in $S$, the jamming Coverage can be defined as the total area that has a value of $\Delta > 1$ [10], so that the presence of the jammers would cause a benefit on the secrecy of the system. Thus, the jamming Coverage can be expressed as

$$\text{Coverage} = \iint_{\Delta>1} dS. \tag{9}$$

On the other hand, the jamming Efficiency is defined as the mean improvement in secrecy over the whole area $S$ [10] and can be expressed as

$$\text{Efficiency} = \frac{1}{|S|}\iint_S \Delta dS. \tag{10}$$

This metric gives a measurement of the average impact of friendly jamming over the secrecy performance. Considering these two metrics and under the observation (after several simulations) that the jamming efficiency value usually oscillates around 1, and doesn't fluctuate much, we can consider the jamming efficiency as a weight factor for the jamming coverage, thus we can define a hybrid metric, the Weighted Secrecy Coverage (WSC), as follows

$$\mathrm{WSC} = \left(\iint_{\Delta>1} dS\right)\left(\frac{1}{|S|}\iint_S \Delta dS\right). \tag{11}$$

This metric considers the area where the secrecy is improved with the presence of friendly jamming joint with the mean improvement over the whole area. The WSC metric is completely determined by $\Delta$, which is obtained for the proposed system as in the following Proposition.

**Proposition 1.** *The jamming improvement, $\Delta$, for the proposed system model, is given by*

$$\Delta = \frac{1 - e^{-\frac{\Omega_{\mathrm{AB}}}{a_{\mathrm{NJ}}}\left(2^{R_S}-1\right)}\left(\frac{1}{2^{R_S}\left(\frac{\Omega_{\mathrm{AB}}}{\Omega_{\mathrm{AE}}}\right)\left(\frac{b_{\mathrm{NJ}}}{a_{\mathrm{NJ}}}\right)+1}\right)}{1 - e^{-\frac{\Omega_{\mathrm{AB}}}{a_{\mathrm{J}}}\left(2^{R_S}-1\right)}\left(\frac{1}{2^{R_S}\left(\frac{\Omega_{\mathrm{AB}}}{\Omega_{\mathrm{AE}}}\right)\left(\frac{b_{\mathrm{J}}}{a_{\mathrm{J}}}\right)+1}\right)}. \tag{12}$$

*Proof:* The proof is provided in Appendix A. ■

## IV. Numerical Results and Discussion

In this Section, we evaluate the impact of different system parameters on the secrecy performance measured by the WSC metric. In the presented results, we aim at narrowing our variable's domains in order to illustrate the overall tendency of our metrics through exhaustive Monte Carlo simulations, as well as validating the derived expressions. The values for environmental and link constants used can be found in [12], and [13] and are $\psi = 9.61$, $\omega = 0.16$, $\alpha = 0.3$, $\alpha_j = 0.3$, $\eta_{\mathrm{NLoS}} = 20$ and $\eta_{\mathrm{LoS}} = 1$ for Urban scenario, which we will be considering. We fix the position of Alice at point $(0, 0)$ without loss of generality and place Bob at distance $d_{AB}$ from Alice along the $x$ axis, at point $(d_{AB}, 0)$, so the cylindrical coordinates of both jammers become $(R_{\mathrm{J}}, \theta_{\mathrm{J1}}, z_{\mathrm{J1}})$ and $(R_{\mathrm{J}}, \theta_{\mathrm{J2}}, z_{\mathrm{J2}})$ respectively with the reference starting from the left of A, $\theta_{\mathrm{J1}}$ opening clockwise and $\theta_{\mathrm{J2}}$ counterclockwise as is illustrated in Fig. 1; and we consider a transmit SNR $\gamma_A = 15$ and jamming SNR $\gamma_{\mathrm{J1}} = \gamma_{\mathrm{J2}} = 15$ if not otherwise specified.

Fig. 2 illustrates the jamming improvement, $\Delta$ versus different position points for E along various radius $r$ of analysis for all angles from 0 to 360° measured counterclockwise starting from the right of Alice. For that purpose we consider three different illustrative cases, (a) with $R_{\mathrm{J}} = 23$, $\theta_{\mathrm{J1}} = \theta_{\mathrm{J2}} = \pi/4$, (b) with $R_{\mathrm{J}} = 14$, $\theta_{\mathrm{J1}} = \theta_{\mathrm{J2}} = \pi/3$ and (c) with $R_{\mathrm{J}} = 7$, $\theta_{\mathrm{J1}} = \theta_{\mathrm{J2}} = \pi/6$. Notice that the obtained graphs show the validation of the theoretical model for various simulation parameters as examples, since they follow precisely the simulation results and that the behaviour is symmetric along 180°, which is to say, behind Alice.

We set the heights of both jammers to be the same, so in Fig. 3 we study the behaviour of the 1 jammer case to determine the optimal height. The Figure shows the WSC obtained with the 1 jammer configuration at cylindrical coordinate $(R_{\mathrm{J}}, \theta_{\mathrm{J}}, z_{\mathrm{J}})$ with a surveillance radius $R_{\mathrm{J}}$ of 7m for all angles $\theta_{\mathrm{J}}$ around Alice, which is to say the radius and angle from Alice to the projection of the jammers in the ground, while $\theta_{\mathrm{J}}$, which lies on the plane, in this case is measured counterclockwise with the usual reference starting from the right hand side of Alice. The optimal height from the ones tried is 13.2m, heights below and above this perform worse. We can also see that for small heights there is positive WSC for every angle around Alice, but for higher UAV heights there are increasing angles for which there is 0 coverage, meaning, it performs as well as, or worse, than the no jammer configuration. Also, it is clear that for any case the maximum performance is obtained at a 180° angle, which is to say, right behind Alice.

Fig. 4 shows the WSC computed for a 2-jammer scenario for different values of the radius around Alice, and the angle points that give the maximum value for our performance metrics are shown per value of $R_{\mathrm{J}}$. The angles shown here are measured from directly behind the line of sight between Alice and Bob and show the angle points for both jammers. We can see that the tendency of having optimal performance for symmetric angles around and behind Alice for both jammers. For higher values of the orbit radius around A, the optimal jammer angles spread farther from the origin, but seem to plateau under 80°.

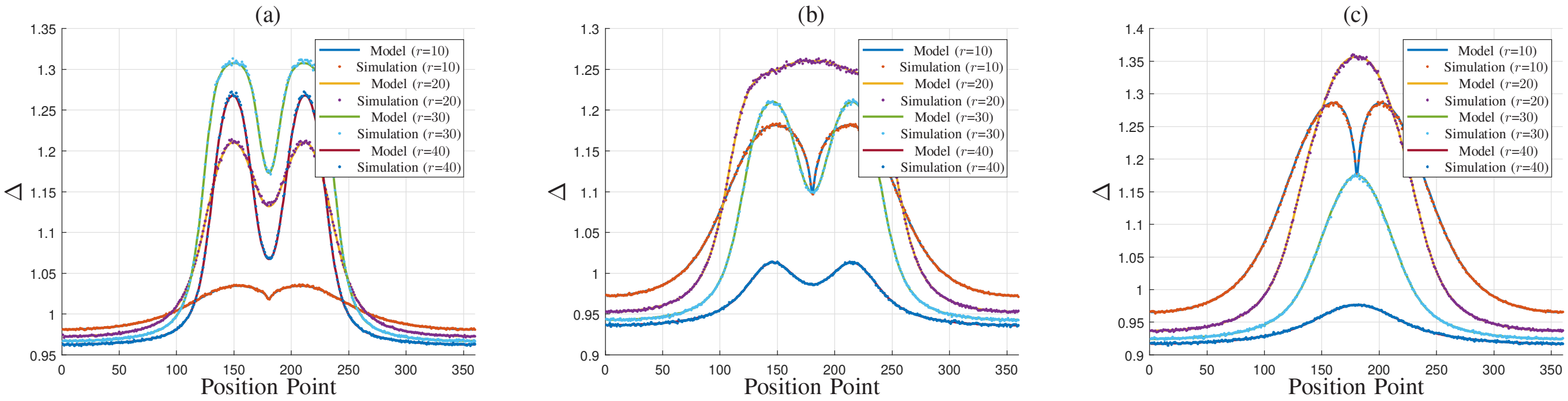

Fig. 2. $\Delta$ versus the position samples for Eve over area $S$. **(a)** $R_{\mathrm{J}} = 23$, $\theta_{\mathrm{J1}} = \theta_{\mathrm{J2}} = \pi/4$, **(b)** $R_{\mathrm{J}} = 14$, $\theta_{\mathrm{J1}} = \theta_{\mathrm{J2}} = \pi/3$, **(c)** $R_{\mathrm{J}} = 7$, $\theta_{\mathrm{J1}} = \theta_{\mathrm{J2}} = \pi/6$

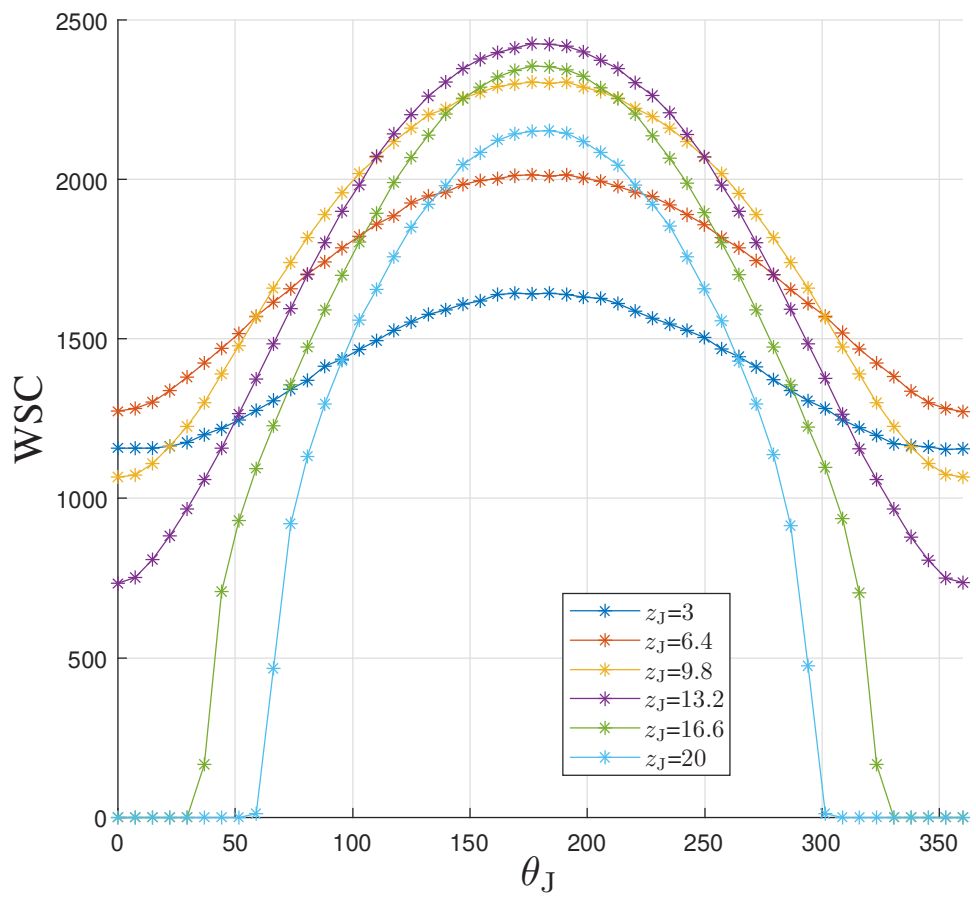

Fig. 3. WSC measured by varying values of jammer height as a function of the jammer angular position around Alice in 1-jammer scenario ($R_{\mathrm{J}} = 7$, $d_{AB} = 20$, $\gamma_A = 15$, $\gamma_{\mathrm{J}} = 15$).

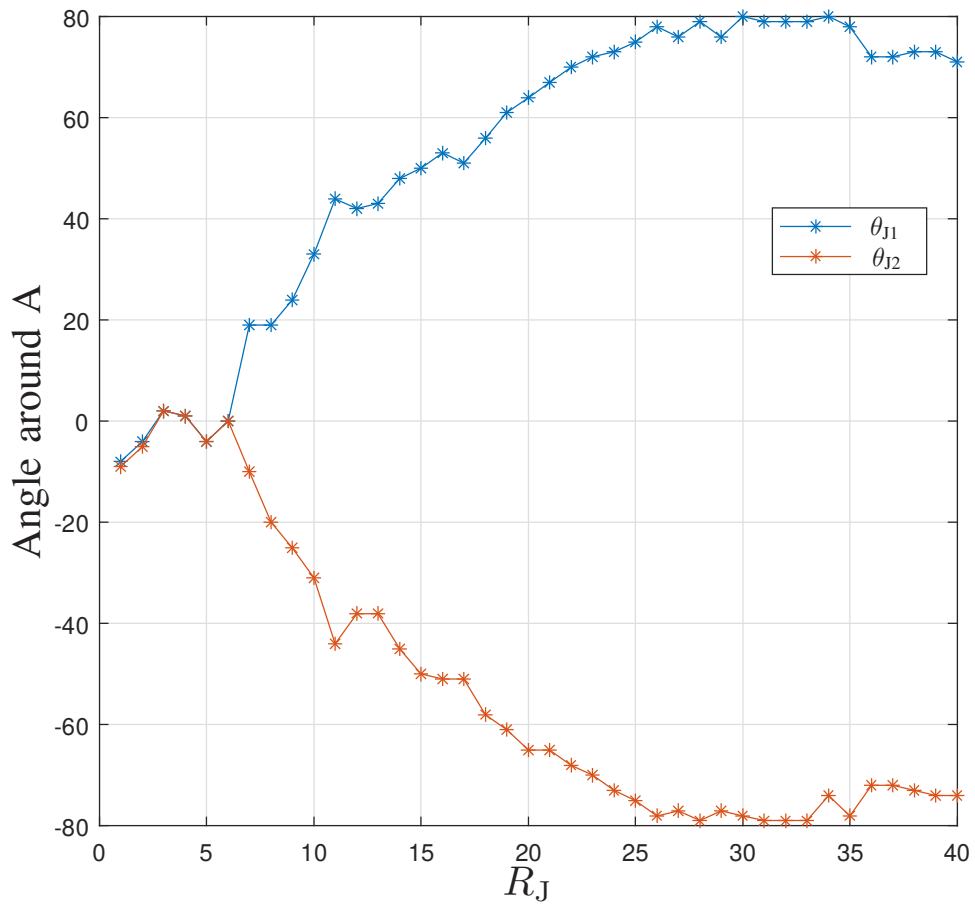

Fig. 4. Angles for both jammers around Alice for which maximum WSC is achieved measured as a function of the radius of the jammers orbit around Alice ($d_{AB} = 20$, $\gamma_A = 15$, $\gamma_{\mathrm{J1}} = \gamma_{\mathrm{J2}} = 15$).

In Fig. 5, we illustrate the WSC obtained by different ratios of jamming SNR allocation between the two jammers by setting the total power available to the legitimate system, where Alice has half of the power allocated, and the other half of the power is allocated between the two jammers. The ratio between the power of both jammers is varied between 0 (all the jamming available power allocated on one of the jammers) and 1 (jamming available power allocated evenly between the jammers) and the WSC computed. It can be seen that the lowest performance values are obtained for the case where all of the power is allocated on a single jammer, and it goes up as the power gets distributed more and more evenly between both of the jammers, which occurs for different values of total available jamming SNR $\gamma_T$. It is clear that the tendency is to find the maximum value for the WSC at 1, which is to say, we have the best power allocation between the jammers when the available jamming SNR is allocated evenly between them.

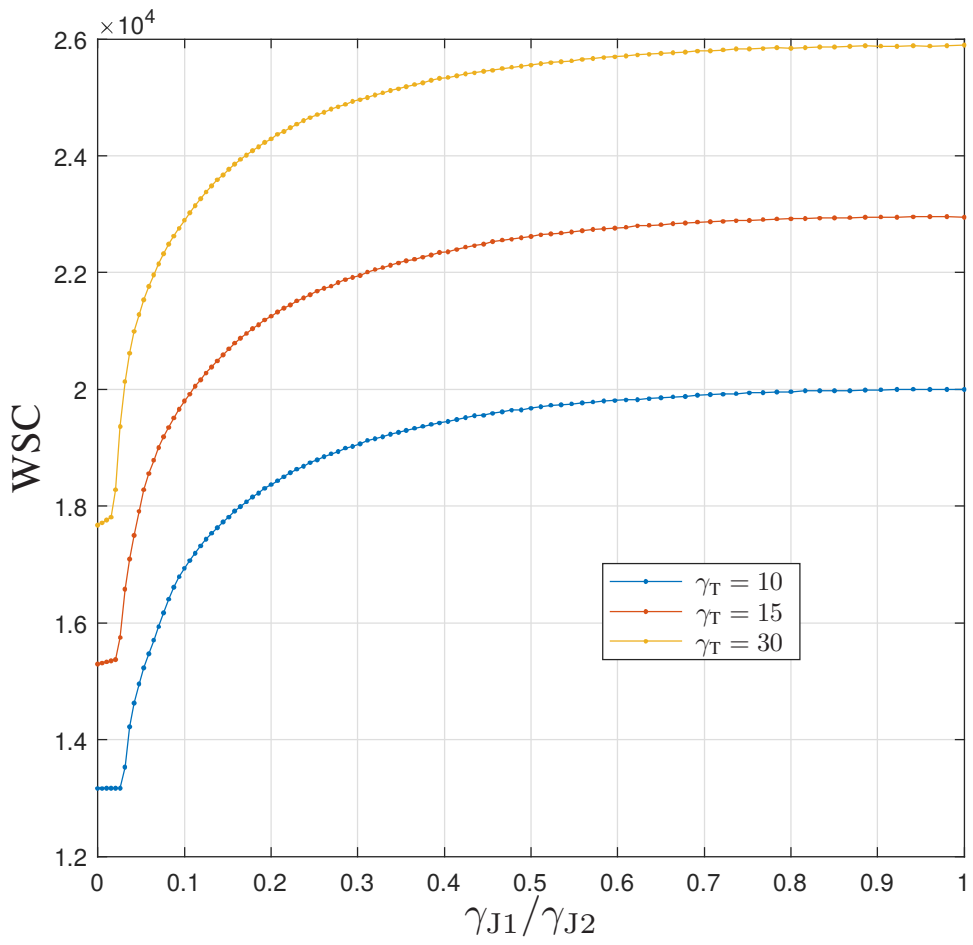

Fig. 5. WSC achieved as a function of the ratio between transmit SNR of both jammers for different values of total available transmit SNR $\gamma_T$ ($R_{\mathrm{J}} = 28$, $d_{AB} = 20$, $\gamma_A = \frac{1}{2}\gamma_T$, $\gamma_{\mathrm{J1}} + \gamma_{\mathrm{J2}} = \frac{1}{2}\gamma_T$).

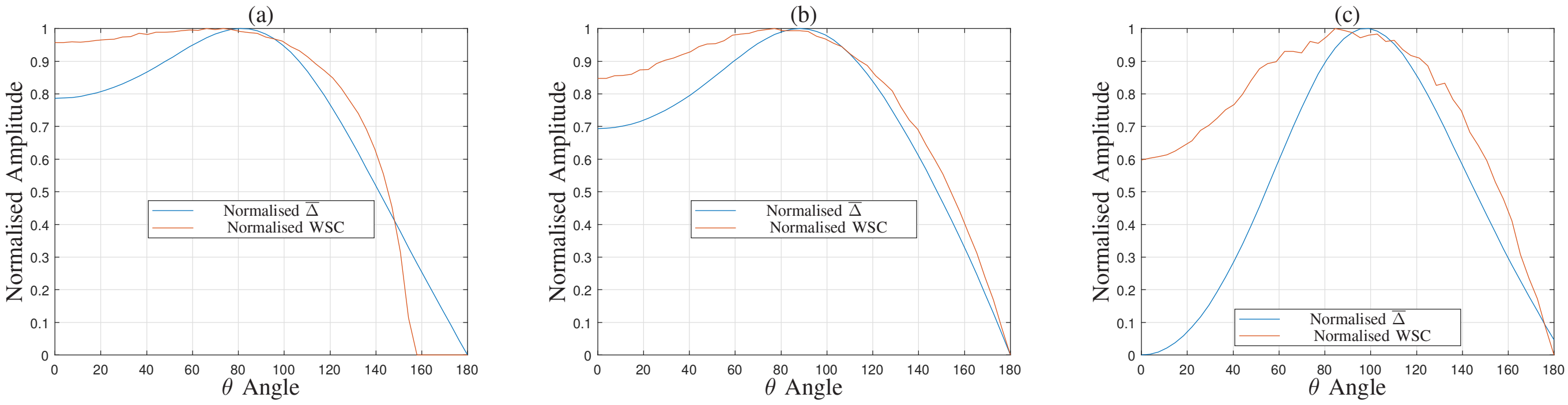

Fig. 6. Behaviour of $\overline{\Delta}$ and Weighted Secrecy Coverage over the whole range of $\theta$ for $z_{\mathrm{J}} = 13$, $\gamma_A = 20$, $\gamma_{\mathrm{J1}} = \gamma_{\mathrm{J2}} = 5$, $R_{\mathrm{J}} = 14$. **(a)** $d_{AB} = 4$, **(b)** $d_{AB} = 14$, **(c)** $d_{AB} = 39$.

### *A. Weighted Secrecy Coverage Maximization*

We fix the surveillance radius of the jammers at a set $R_{\mathrm{J}}$ value and their heights at a given $z_{\mathrm{J}}$ value, and since they are optimally positioned to the back of Alice with the same angle ($\theta_{\mathrm{J1}} = \theta_{\mathrm{J2}}$) from the x axis, we may talk of the opening angle $\theta = \theta_{\mathrm{J1}} + \theta_{\mathrm{J2}}$ as the angle between both jammers in the plane as the only degree of freedom involving the jammers movement. We fix the transmit power of Alice and allocate the available jamming power evenly between both jammers. Then, we have WSC $= \mathrm{WSC}(d_{AB}, \theta)$.

We seek to optimise the Weighted Secrecy Coverage such that the jammers can choose an appropriate angle $\theta$ given a distance $d_{AB}$. For ease of calculations we present an analogous metric $\overline{\Delta}$ defined as

$$\overline{\Delta} = \frac{1 - SOP_{\mathrm{J}}}{1 - SOP_{\mathrm{NJ}}}. \tag{13}$$

The terms $1 - SOP$ refer to the probability of achieving secrecy, which we want to maximise, so $\overline{\Delta}$ has a behaviour equal in nature to $\Delta$, which is, it grows if the presence of the jammers improves secrecy and drops if they impact secrecy negatively. It is easy to show that $\overline{\Delta}$ and $\Delta$ reach 1 at the same points, so using $\overline{\Delta}$ does not impact on the Coverage, only on the Efficiency. Given $\Delta$, straightforward calculations lead to the expression for $\overline{\Delta}$ as:

$$\overline{\Delta} = e^{\Omega_{\mathrm{AB}}\left(2^{R_S}-1\right)\left(\frac{1}{a_{\mathrm{J}}}-\frac{1}{a_{\mathrm{NJ}}}\right)} \left( \frac{2^{R_S}\left(\frac{\Omega_{\mathrm{AB}}}{\Omega_{\mathrm{AE}}}\right)+1}{2^{R_S}\left(\frac{\Omega_{\mathrm{AB}}}{\Omega_{\mathrm{AE}}}\right)\left(\frac{b_{\mathrm{J}}}{a_{\mathrm{J}}}\right)+1} \right), \tag{14}$$

which also accounts for the fact that $a_{\mathrm{NJ}} = b_{\mathrm{NJ}}$. We employ $\overline{\Delta}$ instead of $\Delta$ in the definition of WSC in what follows.

In order to simplify this problem, we seek to find a point where the value of $\overline{\Delta}$ reflects the WSC metric so we optimise over $\overline{\Delta}$ instead; this approach is reasonable, since the WSC is completely determined by $\overline{\Delta}$. We want to show through simulation that $\overline{\Delta}(x_E, y_E)$ at a given $(x_E, y_E)$ computed for a given distance $d_{AB}$ shows the same monotonic behaviour and location of maximum and minimum points than the WSC for the whole range of $\theta$ values. After trying for a number of position and combination of positions, we choose to compute $\overline{\Delta}$ at the following points in the jammers orbit: $(x_A, y_A + R_{\mathrm{J}})$, $(x_A, y_A - R_{\mathrm{J}})$, $(x_A - R_{\mathrm{J}}, y_A)$, and sum each of them, for a given range of $R_{\mathrm{J}}$ orbit radius values, for which we obtain the same monotonic behaviour with the extreme points slightly dislocated for some values of $d_{AB}$, which is the behaviour we were seeking. An example is shown in Fig. 6 where normalised graphs of $\overline{\Delta}$ and of WSC are shown for the sake of monotonic comparison for three different values of $d_{AB}$. We choose the parameters given previous results, namely an equal allocation of jamming power between both jammers and the tried optimal height. Then, for distances $d_{AB}$ higher than 5, the maximum displacement is within 10°. Then, we accept and consider the sub-optimal performance function

$$f(\theta) = \overline{\Delta}(x_A, y_A + R_{\mathrm{J}}) + \overline{\Delta}(x_A, y_A - R_{\mathrm{J}}) + \overline{\Delta}(x_A - R_{\mathrm{J}}, y_A). \tag{15}$$

We then consider our optimisation problem as the maximisation of the univariate function $f(\theta)$ to find the optimal $\theta_{opt}$ which would be expressed in terms of $d_{AB}$. The optimisation problem is then

$$\max_{\theta} \quad f(\theta) \tag{16a}$$

$$\text{subject to} \quad 0 \leq \theta \leq \pi, \tag{16b}$$

We ran this optimisation problem numerically by exhaustive search in a given range of $d_{AB}$ values, for varying $R_{\mathrm{J}}$ radius orbits. Results for this optimisation problem are shown in Fig. 7.

## V. Conclusions

In order to maximise security area metrics, the best positioning for two cooperative jamming UAVs confined to a surveillance orbit around the transmitter, is symmetrical to the line of sight between the legitimate agents of the communication (Alice and Bob) and behind the transmitter. The preferred power allocation between the UAVs is an even split of the total jamming power available between them. Our formulation of the Weighted Secrecy Coverage metric provides a useful concept for secrecy analysis of a system; due to its mathematical complexity approximations have to be used in order to make the optimisation problem tractable. This work ultimately proposed an optimisation problem for sub-optimal secrecy improving positioning of two jamming UAVs in an orbit around Alice for fixed UAV height, power allocation and surveillance radius, as a function of the distance between Alice and Bob and showed results for particular ranges of surveillance radius. Further variable dependency may be considered to obtain a more general expression.

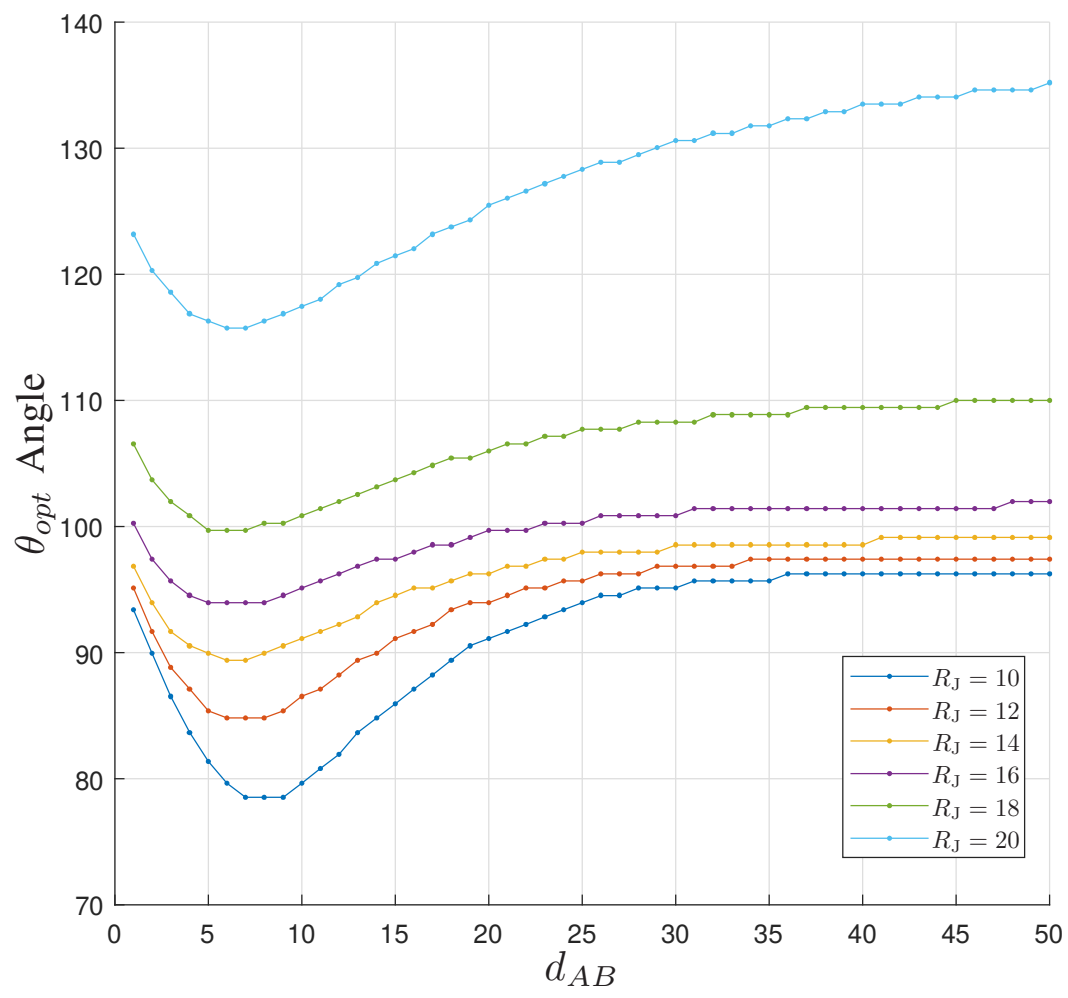

Fig. 7. Optimal aperture angles $\theta$ as a function of distance between A and B ($d_{AB}$) for different orbit radius around Alice $R_{\mathrm{J}}$ for $z_{\mathrm{J}} = 13$, $\gamma_A = 20$, $\gamma_{\mathrm{J1}} = \gamma_{\mathrm{J2}} = 5$.

## VI. Acknowledgements

This work was partially supported by the Academy of Finland, 6Genesis Flagship under Grant 318927 and FAITH project under Grant 334280.

## Appendix A
## Jamming Improvement Derivation

Considering $R_S > 0$, the SOP can be written as follows:

$$\begin{aligned}\mathcal{SOP} &= \Pr\left[\log_2\left(\frac{1+\gamma_B}{1+\gamma_E}\right) < R_S\right] \\ &= \Pr\left[\log_2\left(\frac{1+a\cdot|h_{\mathrm{AB}}|^2}{1+b\cdot|h_{\mathrm{AE}}|^2}\right) < R_S\right] \\ &= \Pr\left[|h_{\mathrm{AB}}|^2 < \frac{1}{a}\left(2^{R_S}\left(1+b\cdot|h_{\mathrm{AE}}|^2\right)-1\right)\right] \end{aligned} \quad (17)$$

This probability is nothing more than the CDF of $|h_{\mathrm{AB}}|^2$ averaged over $|h_{\mathrm{AE}}|^2$, both exponential random variables with parameter $\Omega_{\mathrm{AB}}$ and $\Omega_{\mathrm{AE}}$ respectively, for which it follows:

$$\begin{aligned}\mathrm{SOP} &= \mathrm{E}_{|h_{\mathrm{AE}}|^2}\left[F_{|h_{\mathrm{AB}}|^2}\left(\frac{1}{a}\left(2^{R_S}\left(1+b\cdot|h_{\mathrm{AE}}|^2\right)-1\right)\right)\right] \\ &= \mathrm{E}_{|h_{\mathrm{AE}}|^2}\left[1-e^{-\Omega_{\mathrm{AB}}\frac{1}{a}\left(2^{R_S}(1+b\cdot x)-1\right)}\right] \\ &= 1-\int_0^{\infty} e^{-\Omega_{\mathrm{AB}}\frac{1}{a}\left(2^{R_S}(1+b\cdot x)-1\right)} f_{|h_{\mathrm{AE}}|^2}(x)dx \\ &= 1-\int_0^{\infty} e^{-\Omega_{\mathrm{AB}}\frac{1}{a}\left(2^{R_S}(1+b\cdot x)-1\right)} \Omega_{\mathrm{AE}} e^{-x\Omega_{\mathrm{AE}}}dx \end{aligned}$$

After basic derivations, the expression for the SOP is given by:

$$\mathrm{SOP} = 1 - e^{-\frac{\Omega_{\mathrm{AB}}}{a}\left(2^{R_S}-1\right)}\left(\frac{1}{2^{R_S}\left(\frac{\Omega_{\mathrm{AB}}}{\Omega_{\mathrm{AE}}}\right)\left(\frac{b}{a}\right)+1}\right) \quad (18)$$

So the jamming improvement $\Delta$ is given by:

$$\Delta = \frac{1 - e^{-\frac{\Omega_{\mathrm{AB}}}{a_{\mathrm{NJ}}}\left(2^{R_S}-1\right)}\left(\frac{1}{2^{R_S}\left(\frac{\Omega_{\mathrm{AB}}}{\Omega_{\mathrm{AE}}}\right)\left(\frac{b_{\mathrm{NJ}}}{a_{\mathrm{NJ}}}\right)+1}\right)}{1 - e^{-\frac{\Omega_{\mathrm{AB}}}{a_{\mathrm{J}}}\left(2^{R_S}-1\right)}\left(\frac{1}{2^{R_S}\left(\frac{\Omega_{\mathrm{AB}}}{\Omega_{\mathrm{AE}}}\right)\left(\frac{b_{\mathrm{J}}}{a_{\mathrm{J}}}\right)+1}\right)} \quad (19)$$